\documentclass[12pt]{iopart}

\usepackage{setstack}
\usepackage{iopams}
\usepackage{graphicx}
\usepackage{cite}

\begin{document}

\title{Quantum teleportation with atoms: quantum process tomography}

\author{M. Riebe$^1$, M. Chwalla$^1$, J. Benhelm$^1$, H. H\"affner$^{1,2}$,
W. H\"ansel$^1$, C. F. Roos$^{1,2}$ and R. Blatt$^{1,2}$}

\address{$^1$Institut f\"ur Experimentalphysik, Universit\"at Innsbruck, Technikerstrasse
25, A-6020 Innsbruck}
\address{$^2$Institut f\"ur Quantenoptik und Quanteninformation der
\"Osterreichischen Akademie der Wissenschaften, Technikerstr. 21a, A-6020 Innsbruck}

\begin{abstract}
The performance of a quantum teleportation algorithm implemented on an ion trap quantum computer is
investigated. First the algorithm is analyzed in terms of the teleportation fidelity of six input
states evenly distributed over the Bloch sphere. Furthermore, a quantum process tomography of the
teleportation algorithm is carried out which provides almost complete knowledge about the
algorithm.
\end{abstract}

\section{Introduction}
Quantum teleportation \cite{Bennett93} is one of the fundamental experiments of quantum information
science. The transfer of the quantum properties of one system to a second (distant) system based on
the nonlocal properties of an entangled state highlights the most peculiar and fascinating aspects
of quantum mechanics. The experimental realization of teleportation requires complete experimental
control over a system's quantum state. For this reason, teleportation has been only implemented in
a few physical systems
\cite{Bouwmeester97,Boschi98,Furusawa98,Nielsen98,Pan01,Marcikic03,Riebe04,Barrett04}. One of these
systems are strings of cold ions stored in linear Paul traps. The achievable level of control over
the quantum state of trapped ions makes this system an ideal candidate for quantum information
processing. Single and two-qubit gates constituting the fundamental building blocks for quantum
information processing have already been demonstrated
\cite{Leibfried03,Schmidt-Kaler03,Haljan05,Home06} and characterized by quantum process tomography
\cite{Riebe06}. The concatenation of quantum gates in combination with measurements has been used
for demonstrating simple quantum algorithms \cite{Gulde03,Roos04,Reichle06}. Quantum teleportation
can be viewed as an algorithm that maps one ion's quantum state to another ion. In the context of
quantum communication, teleportation can also be interpreted as a non-trivial implementation of the
trivial quantum channel representing the identity operation. In this paper, we characterize an ion
trap based experimental implementation of such a quantum channel by quantum process tomography. We
improve the previously reported fidelity of the teleportation operation \cite{Riebe04} and extend
the analysis by teleporting the six eigenstates of the Pauli operators $\sigma_{x,y,z}$ and
measuring the resulting density matrices. These data are used for reconstructing the completely
positive map characterizing the quantum channel.

\section{Teleporting an unknown quantum state}
\label{sec:teleportationtheory} Teleportation achieves the
faithful transfer of the state of a single quantum bit between two
parties, usually named Alice and Bob, by employing a pair of
qubits prepared in a Bell state shared between the two parties.
The protocol devised by Bennett et al. \cite{Bennett93} assumes
Alice to be in possession of a quantum state
$\psi_{in}=\alpha|0\rangle+\beta|1\rangle$, where $\alpha$ and
$\beta$ are unknown to Alice. In addition, she and Bob share a
Bell state given by
\begin{equation}
|\Psi_+\rangle=\frac{1}{\sqrt{2}}\left(|0\rangle_A|1\rangle_B+|1\rangle_A|0\rangle_B\right),
\end{equation}
where the subscripts indicate whether the qubit is located in
Alice's or Bob's subsystem. The joint three qubit quantum state of
Alice's and Bob's subsystem
\begin{equation}
|\Psi\rangle_{AB}=\frac{1}{\sqrt{2}}\left(\alpha|00\rangle_A|1\rangle_B+\beta|10\rangle_A|1\rangle_B+\alpha|01\rangle_A|0\rangle_B+\beta|11\rangle_A|0\rangle_B\right)
\end{equation}
can be rearranged by expressing the qubits on Alice's side in
terms of the Bell states
$\Psi^{\pm}=(|10\rangle\pm|01\rangle)/\sqrt{2}$ and
$\Phi^{\pm}=(|00\rangle\pm|11\rangle)/\sqrt{2}$:
\begin{eqnarray}
\fl
|\Psi\rangle_{AB}=\frac{1}{2}(\Phi^+_A\underbrace{\left(\alpha|1\rangle+\beta|0\rangle\right)_B}_{\sigma_x\cdot\Psi_{in}}
+\Phi^-_A\underbrace{\left(\alpha|1\rangle-\beta|0\rangle\right)_B}_{\sigma_z\cdot\sigma_x\cdot\Psi_{in}}
+\Psi^+_A\underbrace{\left(\alpha|0\rangle+\beta|1\rangle\right)_B}_{\Psi_{in}}\nonumber\\
+\Psi^-_A\underbrace{\left(\beta|1\rangle-\alpha|0\rangle\right)_B}_{-\sigma_z\cdot\Psi_{in}})
.
\end{eqnarray}
By a measurement in the Bell basis, Alice projects Bob's qubit
into the states
$\sigma_x\cdot\Psi_{in}$, $(\sigma_x\,\sigma_z)\cdot\Psi_{in}$,
$\Psi_{in}$ and $\sigma_z\cdot\Psi_{in}$ depending on the result
of the measurement. If Alice passes  the measurement result on to
Bob, he is able to reconstruct $\Psi_{in}$ by applying the
necessary inverse operation of either $\sigma_x$,
$\sigma_z\sigma_x$, $I$ or $\sigma_z$ to his qubit.

With trapped ions, it is possible to implement teleportation in a
completely deterministic fashion since both the preparation of the
entangled state and the complete Bell measurement followed by
measurement-dependent unitary transformations are deterministic
operations.

\section{Experimental setup}
 In our experimental setup, quantum information is stored in
superpositions of the $S_{1/2}(m=-1/2)$ ground state and the
metastable $D_{5/2}(m=-1/2)$ state of $^{40}$Ca$^+$ ions. The
calcium ions are held in a linear Paul trap where they form a
linear string with an inter-ion distance of about $5\:\mu$m. State
detection is achieved by illuminating the ion string with light at
397 nm resonant with the $S_{1/2}\leftrightarrow
P_{1/2}$-transition and detecting the resonance fluorescence of
the ions with a CCD camera or a photo multiplier tube. Detection
of the presence or absence of resonance fluorescence corresponds
to the cases where an ion has been projected into the $|S\rangle$
or $|D\rangle$-state, respectively. The ion qubits can be
individually manipulated by pulses of a tightly focussed laser
beam exciting the $|S\rangle\leftrightarrow|D\rangle$ quadrupole
transition at a wavelength of 729 nm. The motion of the ions in
the harmonic trap potential are described by normal modes, which
appear as sidebands in the excitation spectrum of the
$S_{1/2}\leftrightarrow D_{5/2}$ transition. For coherent
manipulation, only the quantum state of the axial center of mass
mode at a frequency of $\omega_{COM}=2\pi\times 1.2$ MHz is
relevant. Exciting ions on the corresponding upper or blue
sideband leads to transitions between the quantum states
$|S,n\rangle$ and $|D,n+1\rangle$, where $n$ is the number of
phonons. By employing sideband laser cooling the vibrational mode
is initialized in the ground state $|n=0\rangle$ and can be
precisely controlled by subsequent sideband laser pulses. These
sideband operations, supplemented by single qubit rotations using
the carrier transition, enable us to implement an entangling
two-qubit quantum gate. Further details of the experimental setup
can be found in \cite{Schmidt-Kaler03b}.

\section{Implementing teleportation in an ion trap}
\begin{figure}
\includegraphics[width=\linewidth]{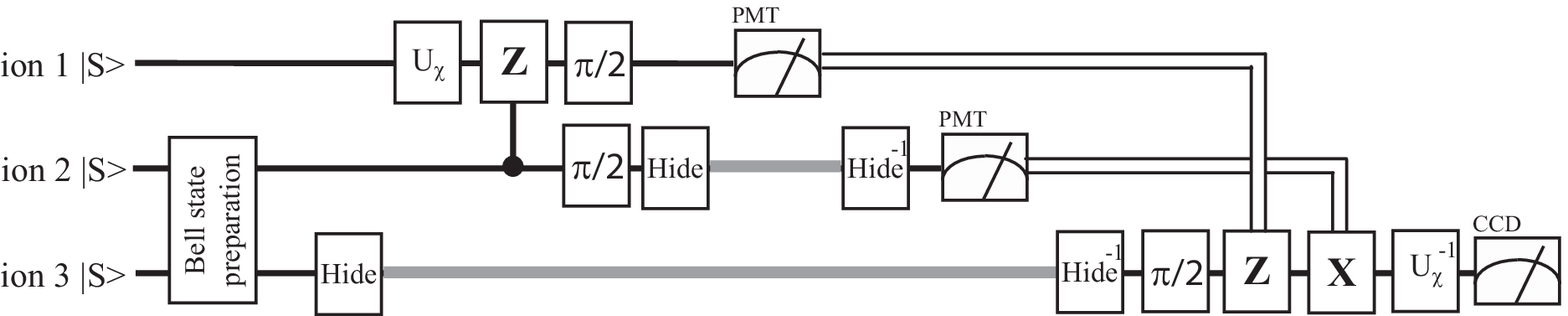}
\caption{Teleportation algorithm for three ion-qubits. Ion 1 is
prepared in the input state $|\chi\rangle=U_{\chi}|S\rangle$ while
ion 2 and 3 are prepared in a Bell state. The teleportation pulse
sequence transfers the quantum information to ion 3. During the
Bell measurement the quantum information in the ions not subjected
to the measurement are protected from the 397 nm light by shifting
the $|S\rangle$-state population to an additional
$|D\rangle$-substate using the pulses denoted by Hide and
Hide$^{-1}$. The operations labelled $X$ and $Z$ represent spin
flip and phase flip operations, respectively.}
\label{fig:telequantumcircuit}
\end{figure}
Three ion-qubits are sufficient for the teleportation experiment.
One qubit carries the unknown quantum information and an entangled
pair of qubits provides the necessary entangled resource for the
information transfer.

Fig.\ \ref{fig:telequantumcircuit} provides an overview of the
pulse sequence used for teleportation. A complete list of all
necessary experimental steps is given in Tab.\
\ref{tab:pulsesequence}. This pulse sequence can be broken down
into the following experimental steps:
\begin{enumerate}
\item {\bf Initialization of ion qubits:} Initially, the ion
string's vibrational motion is laser-cooled by Doppler cooling on
the $S_{1/2}\leftrightarrow P_{1/2}$ dipole transition. Subsequent
sideband cooling on the $S_{1/2}\leftrightarrow D_{5/2}$
quadrupole transition initializes the center-of-mass mode in the
ground state, which is a crucial prerequisite for the entangling
and disentangling sideband operations in the teleportation
circuit. By a pulse of circular polarized 397 nm light, we make
sure that all ion qubits are in the $S_{1/2}(m=-1/2)$ ground state
at the beginning of the teleportation sequence.

\item {\bf Bell state preparation:}  Ion 2 and 3 are prepared in
the Bell state $(|DS\rangle+|SD\rangle)/\sqrt{2}$ by a sequence of
three laser pulses (see Tab.\ \ref{tab:pulsesequence}). We are
able to generate this entangled state with a fidelity of up to
96\% \cite{Roos04b}. Furthermore, this particular Bell state is
highly robust with respect to the major decoherence mechanisms in
our experimental resulting in a lifetime only limited by the
lifetime of the metastable $D_{5/2}$-level \cite{Roos04b}.

\item {\bf Preparation of the input state:} Ion 1 is prepared in
the input state $|\psi_{in}\rangle=U_{\chi}|S\rangle$, where
$U_{\chi}$ is a single qubit rotation.

\item {\bf Rotation into the Bell-basis:} In order to carry out the measurement in the Bell basis,
we have to map the Bell basis onto the product basis $\{|SS\rangle, |SD\rangle, |DS\rangle,
|DD\rangle\}$, which is the natural measurement basis in our setup. This basis transformation is
achieved by first applying a CNOT gate operation to the qubits, mapping the Bell states onto
separable states, and a final Hadamard-like single qubit rotation. In our quantum circuit the CNOT
gate, which is extensively described in \cite{Schmidt-Kaler03b}, is decomposed into a controlled
phase gate and two single qubit rotations of length $\pi/2$. However, one of the $\pi/2$-rotations
(pulse 30 in Tab.\ \ref{tab:pulsesequence}) is shifted to the reconstruction operations on ion 2.
This means that the product basis corresponds to a different set of entangled states, namely
$\{(\Phi^-+\Psi^+)/\sqrt{2}$, $(\Phi^++\Psi^-)/\sqrt{2}$, $(\Phi^+-\Psi^-)/\sqrt{2}$,
$(\Phi^+-\Psi^-)/\sqrt{2}\}$ are mapped onto $\{|DD\rangle$, $|DS\rangle$, $|SD\rangle$,
$|SS\rangle\}$.

\item {\bf Selective read-out of the ion string:} Ion 1 and 2 are
measured in the product basis by illuminating the ions with light
at 397 nm for 250 $\mu$s and detecting the presence or absence of
resonance fluorescence on the $S_{1/2}\leftrightarrow
P_{1/2}$-transition that indicates whether the individual ion was
projected into state $|S\rangle$ or $|D\rangle$. During the
measurement process the coherence of the target ion 3 has to be
preserved. Therefore, the S-state population of ion 3 is
transferred to an additional Zeeman sub-state of the $D_{5/2}$
level, which is not affected by the detection light \cite{Roos04}.
For the detection of the fluorescence light of ion 1 and 2 we use
a photomultiplier (PMT), since its signal can be directly
processed by a digital counter electronics which then decides
which further reconstruction operations are later applied to ion
3. However, this requires to read out the two ions subsequently as
the states $|SD\rangle$ and $|DS\rangle$ cannot be distinguished
with the PMT in a simultaneous measurement of both ions. This is
implemented measuring one ion while hiding the other ion using the
technique described above.

\item {\bf Spin-echo rephasing:} Application of the hiding technique to qubit 3 protects the
quantum information it carries from the influence of the Bell measurement on the other ions.
However, quantum information stored in the D-state manifold is much more susceptible to phase
decoherence from magnetic field fluctuations. In order to undo these phase errors a spin echo
sequence \cite{Hahn50} is applied to qubit 3 (pulse 17 in Tab. \ref{tab:pulsesequence}). In order
to let qubit 3 rephase, a waiting time of 300 $\mu s$ is inserted after completion of the Bell
measurement before the reconstruction operations are applied. Simulations of the teleportation
algorithm show that the spin-echo waiting time which maximizes the teleportation fidelity depends
on the chosen input state. Since a maximum mean teleportation fidelity is desired, a spin-echo time
has to be chosen which is the best compromise between the individual fidelities of the input
states. Additionally, we carry out a spin-echo pulse on ion 1 after the phase gate in order to
cancel phase shifts during the gate operation.

\item {\bf Conditional reconstruction operation:} The information gained in step (v) allows us to
apply the proper reconstruction operations for qubit 3. However, compared to the reconstruction
operations found in Sec. \ref{sec:teleportationtheory} the preset single qubit rotations in our
teleportation circuit have to be modified due to the omitted $\pi/2$-rotation in the Bell
measurement and due to the spin echo applied to ion 3 which acts as an additional $-iY$-rotation.
First of all an additional $\pi/2$-rotation is applied to ion 3, making up the rotation missing in
the Bell analysis. Finally, for the four Bell measurement results $\{|DD\rangle$, $|DS\rangle$,
$|SD\rangle$, $|SS\rangle\}$ the single qubit rotations $\{XZ,iX,iZ,I\}$ have to be applied to
qubit 3, i.e.\ a Z-operation has to be applied whenever ion 1 is found in the $|D\rangle$-state and
an X-operation whenever ion 2 is found to be in $|D\rangle$. Note that all these single qubit
rotations and all following analysis pulses are applied with an additional phase $\phi$. This
allows us to take into account systematic phase errors of qubit 3, by maximizing the teleportation
fidelity for one of the input states by adjusting $\phi$ \cite{Riebe04}. This optimum phase $\phi$
is then kept fixed when teleporting any other quantum states.

\end{enumerate}
\begin{table}
\caption{\small Sequence of laser pulses and experimental steps to
implement teleportation. Laser pulses applied to the i-th ion on
carrier transitions are denoted by $R_i^C(\theta,\varphi)$ and
$R_i^H(\theta,\varphi)$ and pulses on the blue sideband transition
by  $R_i^+(\theta,\varphi)$, where $\theta=\Omega t$ is the pulse
area in terms of the Rabi frequency $\Omega$, the pulse length t
and its phase $\varphi$ \cite{Roos04b}. The index $C$ denotes
carrier transitions between the two logical eigenstates, while the
index $H$ labels transitions from the $S_{1/2}$-- to the
additional $D_{5/2}$--Zeeman substate used to hide individual ion
qubits.} \label{tab:pulsesequence} {\begin{tabular}{|cr|l|l|}
\hline
& & \mbox{Action} & \mbox{Comment} \\
\hline
& 1 & \mbox{Light at 397 nm} & \mbox{Doppler preparation}\\
& 2 & \mbox{Light at 729 nm} & \mbox{Sideband cooling}\\
 & 3 & \mbox{Light at 397 nm} & \mbox{Optical pumping}\\
\hline & 4 & $R^{+}_{3}({\pi}/{2},3 \pi /2)$         & Entangle
ion \#3 with motional
qubit \\
& 5 & $R^{\rm C}_{2}(\pi,3 \pi /2)$ & Prepare ion \#2 for entanglement\\
\raisebox{-1.4ex}[-4.5ex]{\rotatebox{90}{\mbox{Entangle}}} & 6 &
$R^{+}_{2}(\pi,\pi/2)$    & Entangle ion 2 with ion 3  \\
\hline
& 7 &  Wait for 1$\mu$s -- $10\;000$ $\mu$s & Stand--by for teleportation\\
& 8 & $R^{\rm H}_{3}(\pi,0)$         & Hide target ion \\

\hline

& 9 & $R^{\rm C}_{1}(\vartheta_\chi ,\varphi_\chi)$ & Prepare
source ion \#1 in
state $\chi$ \\

\hline
& 10 & $R^{+}_{2}(\pi,3\pi /2)$    & Get motional qubit from ion 2 \\

& 11 & $R^{+}_{1}(\pi/\sqrt{2},\pi/2)$      & Composite pulse for
phasegate
\\
& 12 & $R^{+}_{1}(\pi,0)$               & Composite pulse for phasegate \\
& 13 & $R^{+}_{1}(\pi/\sqrt{2},\pi/2)$      & Composite pulse for
phasegate
\\
& 14 & $R^{+}_{1}(\pi,0)$       & Composite pulse for phasegate \\

& 15 & $R^{\rm C}_{1}(\pi,\pi/2)$   & Spin echo on ion 1 \\

& 16 & $R^{\rm H}_{3}(\pi,\pi)$     & Unhide ion 3 for spin echo \\
& 17 & $R^{\rm C}_{3}(\pi,\pi/2)$   & Spin echo on ion 3 \\
& 18 & $R^{\rm H}_{3}(\pi,0)$       & Hide ion 3 again\\

& 19 & $R^{+}_{2}(\pi,\pi/2)$              & Write motional qubit
back to ion
\#2 \\

& 20 & $R^{\rm C}_{1}(\pi /2,3 \pi/2)$      & Part of rotation
into Bell--basis
\\

\raisebox{4ex}[-7.5ex]{\rotatebox{90}{\mbox{Rotate into
Bell--basis}}} & 21 &
$R^{\rm C}_{2}(\pi /2,\pi /2)$  & Finalize rotation into Bell basis\\

\hline

& 22 & $R^{\rm H}_{2}(\pi,0)$   & Hide ion 2 \\
& 23 & PMT detection \#1 (250 $\mu$s) & Read out ion 1 with photomultiplier\\
& 24 & $R^{\rm H}_{1}(\pi,0)$   & Hide ion 1 \\
& 25 & $R^{\rm H}_{2}(\pi,\pi)$   & Unhide ion 2 \\
& 26 & PMT detection \#2 (250 $\mu$)s & Read out ion 2 with photomultiplier\\
\raisebox{-1.3ex}[-7.5ex]{\rotatebox{90}{\mbox{Read--out}}} & 27 &
$R^{\rm
H}_{2}(\pi,0)$   & Hide ion \#2 \\

& 28 & Wait 300 $\mu$s & Let system rephase; part of spin echo \\

& 29 & $R^{\rm H}_{3}(\pi,\pi)$       & Unhide ion 3\\
\hline
& 30 & $R^{\rm C}_{3}(\pi/2,3 \pi /2+\phi)$ & Change basis\\

\hline

& 31 & $R^{\rm C}_{3}(\pi,\phi)$        & i$\sigma_x$ \\
& 32 & $R^{\rm C}_{3}(\pi,\pi/2+\phi)$   & -i$\sigma_y$ \hspace{-0.3cm}
\raisebox{1.5ex}[0ex]{$\bigg\}\;{=-i\sigma_z}$
\begin{minipage}{4cm}{\begin{flushleft}{conditioned on PMT detection \#1}\end{flushleft}}\end{minipage}} \\
\raisebox{-1.3ex}[-7.5ex]{\rotatebox{90}{\begin{minipage}{3cm}{Recon-
\\struction}\end{minipage}}} & 33 & $R^{\rm C}_{3}(\pi,\phi)$   & i$\sigma_x$
conditioned on PMDetection 2\\

\hline & 34 & $R^{\rm
C}_{3}(\vartheta_\chi,\varphi_\chi+\pi+\phi)$ & Inverse of
preparation of $\chi$ with offset $\phi$ \\
\hline

& 35 & Light at 397 nm& Read out ion 3 with camera \\
\hline

\end{tabular}}
\end{table}

\section{Teleportation results}
Due to experimental imperfections and interaction of the qubits with the environment, no
experimental implementation of teleportation will be perfect. For this reason, we describe the
experimental teleportation operation by a completely positive map $\mathcal{E}(\rho)$, expressed in
operator sum representation as \cite{NielsenChuang00}:
\begin{equation}
\mathcal{E}(\rho)=\sum_{m,n=1}^{4}\chi_{mn}\;A_m\rho
A_n^{\dagger}, \label{eq:operatorsum}
\end{equation}
where $\rho$ is the input state to be teleported, and
$A_m\in\{I,\sigma_x,\sigma_y,\sigma_z\}$ is a set of operators
forming a basis in the space of single-qubit operators. The
process matrix $\chi$ contains all information about the
state-mapping from qubit 1 to qubit 3.

A useful quantity characterizing the quantum process ${\cal E}$ is
the average fidelity $\bar{F}=\int d\psi\langle\psi|{\cal
E}(\psi)|\psi\rangle$ where the average over all pure input states
is performed using a uniform measure on state space with $\int
d\psi=1$. In the case of a single qubit process, the integral
would be over the surface of the Bloch sphere. However, for the
calculation of $\bar{F}$, an average over a suitably chosen finite
set of input states suffices \cite{Bowdrey02, Nielsen02}. Using
the eigenstates $\psi_{\pm k}, k\in\{x,y,z\},$ of the Pauli
matrices $\sigma_x,\sigma_y,\sigma_z$, $\bar{F}$ is obtained by
calculating $\bar{F}=\frac{1}{6}\sum_{j\in\{\pm x,\pm y,\pm z\}}
\langle\psi_j|{\cal E}(\psi_j)|\psi_j\rangle$.

\begin{figure}[ht]
\includegraphics[width=\linewidth]{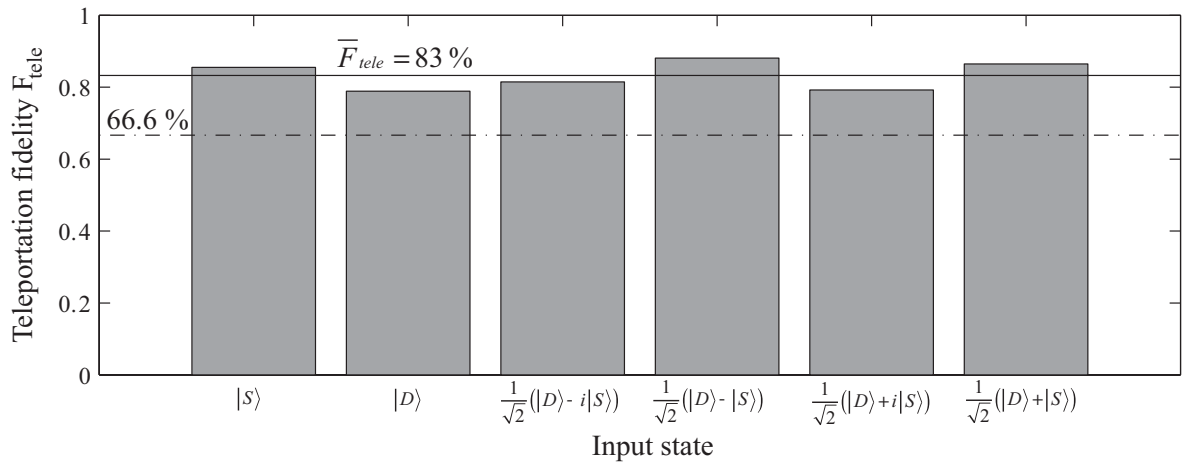}
\caption{Measured teleportation fidelity $F_{tele}$ for six different input states. All fidelities
are well above the 2/3 threshold proving successful quantum teleportation. The average
teleportation fidelity is $\bar{F}_{tele}=83(1)\%$.} \label{fig:telefidelity}
\end{figure}

The overlap $\langle\psi_j|{\cal E}(\psi_j)|\psi_j\rangle$ between the input state $\psi_j$
prepared in ion qubit \#1 with the output state generated via teleportation in ion qubit \#3 is
measured directly in our experiment by applying the inverse unitary transformation to ion qubit \#3
after teleportation and determining the probability to find this qubit in the initial state
$|S\rangle$, i.e.\ formally the teleportation fidelity is given by $F_{tele}=\langle
S|U_{\chi}^{-1}\rho_{exp}U_{\chi}|S\rangle$, where $\rho_{exp}$ is the quantum state of ion qubit
\#3 after teleportation. For the six input states $\psi_1=|S\rangle$, $\psi_2=|D\rangle$,
$\psi_3=(|D\rangle-i|S\rangle)/\sqrt{2}$, $\psi_4=(|D\rangle-|S\rangle)/\sqrt{2}$,
$\psi_5=(|D\rangle+i|S\rangle)/\sqrt{2}$, $\psi_6=(|D\rangle+|S\rangle)/\sqrt{2}$, the
teleportation fidelities range between 79\% and 87\% (see Fig.\ \ref{fig:telefidelity}), with an
average fidelity of $\bar{F}=83(1)\%$. This average fidelity proves successful operation of the
teleportation algorithm, as it exceeds the maximum value of 2/3 that is achievable without using
entangled states \cite{Massar95}.

\begin{figure}[ht]
\includegraphics[width=\linewidth]{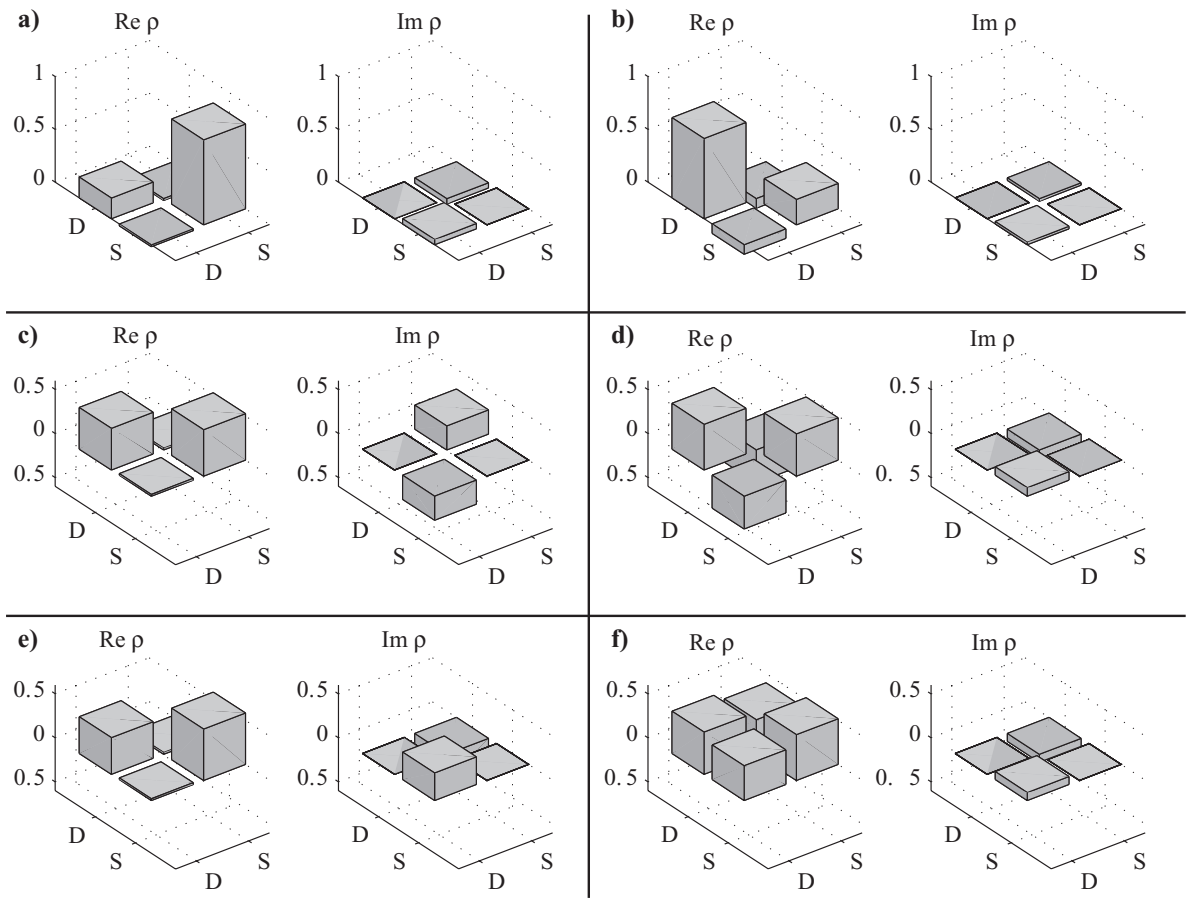}
\caption{Real and imaginary part of the density matrix of the
output qubit for the six different input states a)
$\psi_1=|S\rangle$, b) $\psi_2=|D\rangle$, c)
$\psi_3=(|D\rangle-i|S\rangle)/\sqrt{2}$, d)
$\psi_4=(|D\rangle-|S\rangle)/\sqrt{2}$, e)
$\psi_5=(|D\rangle+i|S\rangle)/\sqrt{2}$, f)
$\psi_6=(|D\rangle+|S\rangle)/\sqrt{2}$.}
\label{fig:teletomografien}
\end{figure}

A more complete way of characterizing the teleportation process is
achieved by determining the output state of qubit 3 by quantum
state tomography, which requires measurements in three different
measurement bases. From these measurements, the density matrix of
the output qubit is estimated using a maximum likelihood algorithm
\cite{Roos04b}. The resulting density matrices of the six input
states are shown in Fig. \ref{fig:teletomografien}.

\begin{figure}
\includegraphics[width=\linewidth]{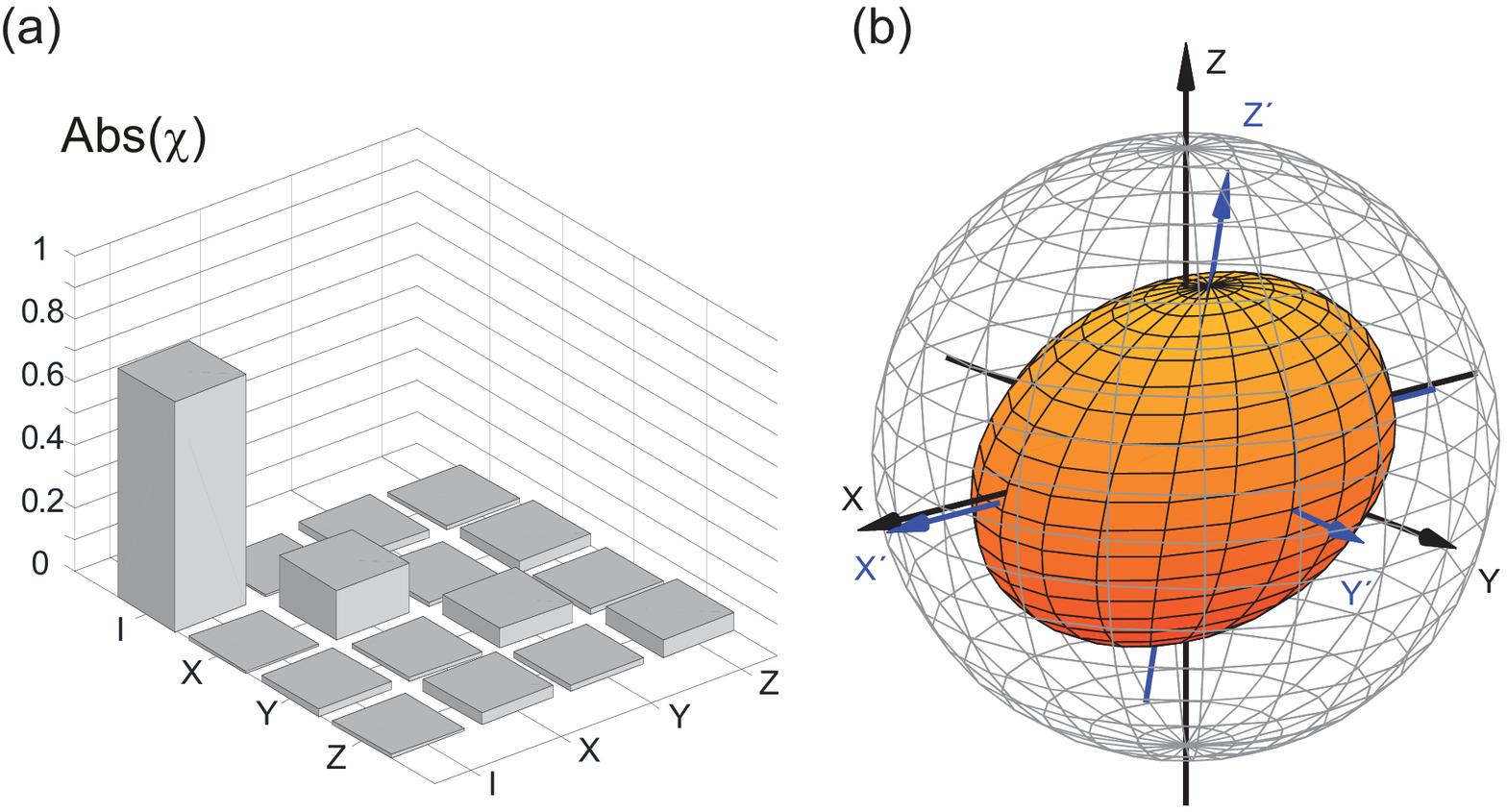}
\caption{Results of process tomography of teleportation algorithm. In (a) the absolute value of
process matrix $\chi$ is shown. The dominating diagonal element is the identity with
$\chi_{II}=0.73(1)$. The plot drawn in (b) shows how the input states lying on the surface of the
initial Bloch sphere (meshed surface) are transformed by the teleportation algorithm, with the
output states lying on the solid surface.} \label{fig:teleproctomresults}
\end{figure}

Full information about the relation between the input and output of the teleportation algorithm is
gained by a quantum process tomography. This procedure requires to determine the output state
$\mathcal{E}(\rho_i)$ after application of the investigated operation for a set of at least four
linear independent input states $\rho_i$. With this data, the process matrix $\chi$ is obtained by
inverting equation (\ref{eq:operatorsum}). Due to inevitable statistical errors in the measurement
process the resulting $\chi$ will in general not be completely positive. This problem is avoided by
employing a maximum likelihood algorithm, which determines the completely positive map which yields
the highest probability of producing the measured data set. We use the tomographically
reconstructed input states $\psi_1$-$\psi_6$ for a determination of the process matrix $\chi$ by
maximum likelihood estimation \cite{Jezek03}. The absolute value of the elements of the resulting
process matrix $\chi_{tele}$ is shown in Fig.\ \ref{fig:teleproctomresults}a). As expected, the
dominant element is the identity with $\chi_{II}=0.73(1)$, which is identical to the process
fidelity $F_{proc}=tr(\chi_{idtele}\chi_{tele})$, where $\chi_{idtele}$ denotes the ideal process
matrix of the teleportation algorithm. This agrees well with the average fidelity stated above, as
average and process fidelity are related by $\bar{F}=(2F_{proc}+1)/3$ for a single qubit map
\cite{Nielsen02}.

A quantum process operating on a single quantum bit can be
conveniently represented geometrically by picturing the
deformation of a Bloch sphere subjected to the quantum process
\cite{NielsenChuang00}. The quantum operation maps the Bloch
sphere into itself by deforming it into an ellipsoid that may be
rotated and displaced with respect to the original sphere
representing the input states. This transformation is described by
an affine map $r_{out}=OSr_{in}+b$ between input and output Bloch
vectors where the matrices $O$ and $S$ are orthogonal and
positive-semidefinite, respectively. Fig.\
\ref{fig:teleproctomresults}b) shows the result for the
teleportation algorithm. The transformed ellipsoid is centered at
$b\approx(0, 0.09, -0.05)$ with errors of about $\pm 0.03$ for
each coordinate. The matrix $S$ shrinks the sphere anisotropically
(its eigenvalues are 0.78, 0.58, 0.55), $O$ rotates the sphere by
an angle of about $2^\circ(2)$. This demonstrates that the loss of
fidelity is mostly due to decoherence and not caused by an
undesired unitary operation rotating the sphere as the orientation
of the deformed Bloch sphere hardly differs from the orientation
of the initial sphere. The results are consistent with the
assumption that the rotation matrix $O$ is equal to the identity
as desired.

\section{Conclusion}
We demonstrated deterministic teleportation of quantum information between two atomic qubits. We
improve the mean teleportation fidelity $\bar{F}=75\%$ reported in \cite{Riebe04} to
$\bar{F}=83(1)\%$ and unambiguously demonstrate the quantum nature of the teleportation operation
by teleporting an unbiased set of six basis states \cite{vanEnk06} and using the data for
completely characterizing the teleportation operation by quantum process tomography. The process
tomography result shows that the main source of infidelity is decoherence while systematic errors
are negligible. To make further progress towards high-fidelity quantum operations, decoherence
rates have to be reduced by either reducing environmental noise or encoding quantum information in
noise-tolerant quantum states \cite{Langer05}.

\end{document}